\documentclass[conference]{IEEEtran}
\usepackage{cite}
\usepackage{amsmath,amssymb,amsfonts}
\usepackage{algorithmic}
\usepackage{graphicx}
\usepackage{textcomp}
\usepackage{caption} 
\usepackage[dvipsnames]{xcolor}

\usepackage{pifont}%
\newcommand{\cmark}{\textcolor{green}{\ding{51}}}%
\newcommand{\xmark}{\textcolor{red}{\ding{55}}}%
\usepackage{colortbl}
\newcommand{\graycell}[1]{\cellcolor{teal!15}#1}

\usepackage{multirow}
\usepackage{arydshln}
\usepackage{tcolorbox}

\usepackage[normalem]{ulem}

\usepackage{graphicx}
\usepackage{etoolbox}
\newcommand{\insertfigteaser}{
\vspace{2mm}
\centering
\captionsetup{type=figure}
\includegraphics[width=1.0\textwidth]{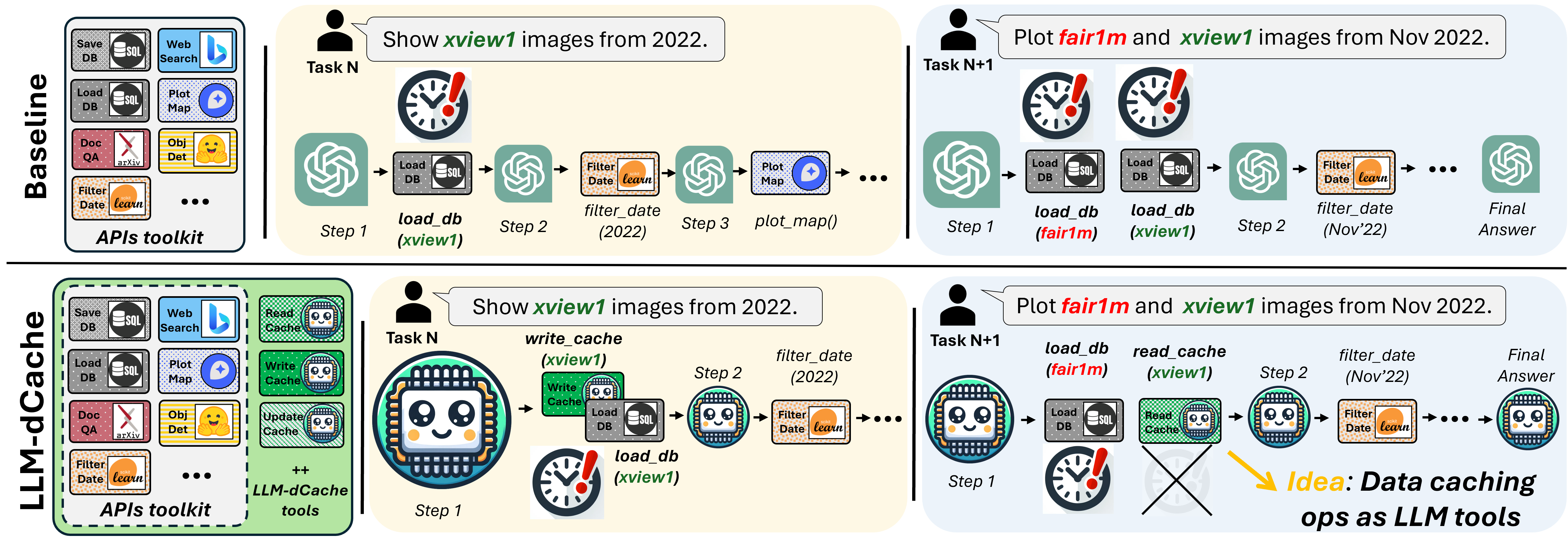}
\captionof{figure}{GPT-driven data caching with LLM-dCache: we expose cache operations as callable API tools to GPT, enabling it to read and update cache data dynamically to respond to user queries. On a large-scale Copilot platform with hundreds of GPTs and terabytes of imagery, LLM-dCache speeds up task-completion by 1.24$\times$ on average across various prompting techniques.}
\label{fig:agent}
}

\makeatletter
\apptocmd{\@maketitle}{\setcounter{figure}{0}\centering\insertfigteaser}{}{}
\makeatother

\def\BibTeX{{\rm B\kern-.05em{\sc i\kern-.025em b}\kern-.08em
    T\kern-.1667em\lower.7ex\hbox{E}\kern-.125emX}}
    
\begin{document}

\title{LLM-dCache: Improving Tool-Augmented LLMs with GPT-Driven Localized Data Caching}

\author{
\IEEEauthorblockN{Simranjit Singh$^{*1}$, Michael Fore$^{*1}$, Andreas Karatzas$^{*2}$, Chaehong Lee$^1$, Yanan Jian$^1$,}
\IEEEauthorblockN{Longfei Shangguan$^3$, Fuxun Yu$^1$, Iraklis Anagnostopoulos$^2$, Dimitrios Stamoulis$^1$}
\IEEEauthorblockA{Email:\{simsingh, mifore, chaelee, yananjian,  fuxunyu, distamo\}@microsoft.com}
\IEEEauthorblockA{longfei@pitt.edu, \{andreas.karatzas, iraklis.anagno\}@siu.edu -- $^*$ Equal contribution}
\IEEEauthorblockA{$^1$Microsoft Corporation, USA~~~$^2$Southern Illinois University, USA~~~$^3$University of Pittsburgh, USA}
}


\maketitle

\begin{abstract}
As Large Language Models (LLMs) broaden their capabilities to manage thousands of API calls, they are confronted with complex data operations across vast datasets with significant overhead to the underlying system. In this work, we introduce LLM-dCache to optimize data accesses by treating cache operations as callable API functions exposed to the tool-augmented agent. We grant LLMs the autonomy to manage cache decisions via prompting, seamlessly integrating with existing function-calling mechanisms. Tested on an industry-scale massively parallel platform that spans hundreds of GPT endpoints and terabytes of imagery, our method improves Copilot times by an average of 1.24$\times$ across various LLMs and prompting techniques.
\end{abstract}



\section{Introduction}

Recent advances in Large Language Models (LLMs) have enhanced their reasoning capabilities towards solving complex problems, allowing them to manage thousands of tools and API calls efficiently~\cite{zhuang2024toolqa, qin2024toolllm}. These improvements have unlocked their potential across large-scale systems, including UI/Web interfaces~\cite{singh2024evaluating,zhou2023webarena}, mobile apps~\cite{koh2024visualwebarena}, SQL backends~\cite{guo2023retrievalaugmented}, and remote sensing platforms~\cite{singh2024geoqa}. These uses exemplify system-level complexity by requiring integration of various APIs for loading, filtering, processing, and visualizing data across multiple temporal and spatial dimensions~\cite{yuan2024llm}. 

As Copilots scale, the overhead on the underlying stack increases, from cloud endpoints to local execution devices~\cite{hu2024routerbench,fore2024geckopt}, catalyzing a fundamental shift in how we design large-scale LLM-based systems and software~\cite{kim2024llmcompiler, zhou2023webarena}. However, early system optimizations primarily target simplified queries or well-defined benchmarks~\cite{ning2023skeleton} that might not capture nuanced task patterns and data dependencies at the system level~\cite{singh2024geollmengine}. In realistic LLM workloads, data exhibits significant reusability. Consider an analyst who asks ``\textit{show me satellite images around Newport Beach, CA.}'' with a subsequent prompt ``\textit{Now, detect airplanes in \textbf{this} area},'' demonstrating a scenario where data elements are repeatedly accessed. 

In this work, we draw inspiration from spatiotemporal reusability patterns akin to those observed in CPU cache systems and we introduce LLM-dCache, a GPT-driven caching strategy to optimize LLM data access patterns. Our \textbf{key intuition} lies in a novel design choice to seamlessly integrate cache management as one of the LLM tools, facilitating a fully GPT-driven \textit{plug-and-play} approach compatible with existing function-calling mechanisms with minimal overhead. Evaluated on a large-scale geospatial platform~\cite{singh2024geollmengine}, LLM-dCache achieves latency reductions across various agents. We hope these findings motivate further exploration into empowering LLMs with other system level optimizations.

\section{Related work}

\textbf{Model-level LLM optimization}: several works aim to enhance LLM efficiency via model design improvements, such as quantization~\cite{dettmers2024qlora}, pruning~\cite{ma2023llmpruner}, KV token caching~\cite{kwon2023kv}, or token compression~\cite{jiang2023llmlingua}. Despite these advances, as motivated in~\cite{kim2024llmcompiler}, these techniques might have limitation in scenarios involving immutable black-box LLM models within cloud-based systems, where direct modifications to models and their inference mechanisms are limited~\cite{fore2024unlearning}. We therefore focus on design optimizations at the system level~\cite{packer2024memgpt,singh2024llmcompiler}, which are especially important in large-scale Copilot platforms.

\textbf{Application-level LLM optimization}: methodologies such as MemGPT~\cite{packer2024memgpt} and ``model-as-a-resource'' caching~\cite{xu2024cached} align with our motivation. We also note advancements from the open-source community, with LangChain now supporting prompt-caching~\cite{langchaincache}. Similarly, drawing from hardware design and parallel computing, recent methods~\cite{kim2024llmcompiler, ning2023skeleton, singh2024llmcompiler} explore parallel execution strategies. While these methods offer benefits for parallel or repeating tasks, they overlook the critical aspect of data locality, as they assume task chains with short-horizons~\cite{kim2024llmcompiler} or template-based question-answer pairs with simplistic task interdependencies~\cite{ning2023skeleton}.

\section{Methodology}

Our goal is to design and assess LLM-dCache on realistic data patterns in large-scale cloud-first Copilot systems. 

\textbf{Cache operations}: We aim to explore GPT's ability to understand \textit{when} to read and use caching to execute a given task, as well as whether GPT is able to effectively implement a cache update policy autonomously. To allow GPT to handle system-level decisions via in-context prompting, we therefore define the operation of loading cache data as a tool in \textit{GPT function calling}, \textit{i.e.}, exposing its function definition in the GPT API call alongside other tool descriptions. Upon receiving a user query, GPT is informed of the current cache contents and decides whether to execute the cache loading tool. 

Similarly, we experiment with an entirely prompt-based implementation of cache updating. We succinctly describe the update policy to GPT and furnish it with this round's load operations and cache contents in JSON format, then query GPT to return the updated cache state. We opt for the Least Recently Used (LRU) scheme as our primary cache update strategy, while we ablate other schemes.

Framing caching functions as GPT tools streamlines our implementation and makes it platform-agnostic. The cache read and update operations become part of GPT's decision-making process, thus requiring minimal changes. Additionally, granting the LLM autonomy over cache decisions allows our method to handle cache misses: upon a failed function call, the LLM is prompted to reassess its tool sequence, just as it would any other tool-selection missteps where the API return-message indicates a failure. This abstraction, simulating a main memory read after a cache-hit scenario and managed entirely at the LLM level, effectively positions the LLM as a memory controller. Such dynamic adaptability is key to rectifying inaccuracies in tool selection in real-time.

\begin{tcolorbox}[title=LLM-dCache prompting, colback=gray!20, colframe=gray!75, rounded corners, sharp corners=northeast, sharp corners=southwest]
\footnotesize
\texttt{As a Compiler handling geospatial data, you have access to the following tools [..] \\
~\\
\textbf{Tools}: \\
- load\_db(..images from  database..)\\
- read\_cache(..images from local cache..)\\
- ...\\
~\\
Given the user query, the cache content, and the examples below, complete the task [..]  \\
~\\
\textbf{User Query}: \{question\}\\
\textbf{Cache}: \{cache content\}\\
~\\
------\\
\textbf{Example 1}: \\
Query: Plot the xview1 images from 2022\\
Cache: \{,\}\\
Thought: The user asks for the \textcolor{ForestGreen}{\textit{\textbf{xview1-2022}}} imagery. The cache is empty [..]\\
Action: To complete the task I will call load\_db(\textcolor{ForestGreen}{\textit{\textbf{xview1-2022}}}), then [..]~\\
Answer: .. \\
~\\
------\\
~\\
\textbf{Example 2}: \\
Query: Show fair1m and xview1 imgs from 2022\\
Cache: \{\textcolor{ForestGreen}{\textit{\textbf{xview1-2022}}},\}\\
Thought: The user wants both the \textcolor{red}{\textit{\textbf{fair1m-2022}}} and \textcolor{ForestGreen}{\textit{\textbf{xview1-2022}}} images. The cache is already contains the latter, so [..]\\
Action: To complete the task I will first call load\_db(\textcolor{red}{\textit{\textbf{fair1m-2022}}}), then read\_cache(\textcolor{ForestGreen}{\textit{\textbf{xview1-2022}}}), and [..]~\\
Answer: .. 
}
\end{tcolorbox}

\textbf{Cache specifications}: We represent and retrieve data as key-value pairs. As we operate on top of geospatial data, we opt to use the string template \textit{dataset}-\textit{year} as cache keys. We find this temporal granularity to be the most sensible (as opposed to longitude-latitude coordinates due to the spatial skewness of data around regions of interest like major cities). We then store as values the GeoPandas DataFrames containing the respective yearly imagery metadata -- filenames, coordinates, detections, timestamps, \textit{etc}. As is common in many geospatial platforms, the actual image files are not loaded into memory until needed for specific subsequent operations. As the yearly GeoPandas DataFrames typically occupy 50-100 MB, so we find it reasonable to set a cache size limit of 5 entries at a time. We note that such design choices are likely to be application specific, and we leave further ablations for future work. 

\begin{table*}[htbp]
\caption{LLM-dCache achieves latency reductions across models and prompting techniques with no degradation in overall agentic performance, as agent metrics are within established variance bounds~\cite{singh2024llmcompiler}.}
\begin{center}
\begin{tabular}{|c|c|c|c|c|c|c|c|c|c|c|}
\hline
\multirow{2}{*}{Model} & \multirow{2}{*}{LLM-dCache} & Success  & Correctness  & Obj. Det & LCC  & VQA  & Avg Tokens & Avg Time & \multirow{2}{*}{Speedup $\uparrow$}  \\
 &  & Rate (\%) $\uparrow$ & Rate (\%) $\uparrow$ & F1 (\%) $\uparrow$ & R (\%) $\uparrow$ & Rouge-L $\uparrow$ & / Task $\downarrow$ & / Task (s) $\downarrow$ & \\
\hline
\multicolumn{10}{|l|}{\textbf{~GPT-3.5 Turbo}} \\ \hline
\multirow{2}{*}{CoT - Zero-Shot} & \xmark & 49.45 & 38.47 &  70.68 & 70.19 & 56.62 & 25.23k & 6.96 & -- \\
& \graycell\cmark & \graycell 49.40 & \graycell 37.96  & \graycell 69.71 & \graycell 71.23 & \graycell 55.57 & \graycell 25.55k & \graycell 5.67  & \graycell 1.23 $\times$ \\ \hline
\multirow{2}{*}{CoT - Few-Shot}  & \xmark & 54.42 & 70.50 & 89.03 & 82.19 & 62.58 & 30.81k & 6.52 & --  \\
& \graycell \cmark   & \graycell 54.07 & \graycell 69.61  & \graycell 88.12 & \graycell 81.31 & \graycell 62.08 & \graycell 30.02k & \graycell 5.29 & \graycell 1.23 $\times$ \\ \hline
\multirow{2}{*}{ReAct - Zero-Shot}  & \xmark  & 50.85 & 70.04 & 87.94 & 89.12 & 61.41 & 27.09k & 7.29 &  \\
 & \graycell \cmark  & \graycell 50.47 & \graycell 68.91 & \graycell 80.42 & \graycell 89.31 & \graycell 60.78 & \graycell 27.65k & \graycell 5.47 & \graycell 1.33 $\times$ \\ \hline
\multirow{2}{*}{ReAct - Few-Shot} &  \xmark & 63.45 & 71.06 & 82.59 & 92.36 & 69.35 & 34.40k & 6.64 & --  \\
 & \graycell \cmark  & \graycell 63.14 & \graycell 69.17 & \graycell 81.19 & \graycell 88.41 & \graycell 65.76 & \graycell 34.86k & \graycell 5.77 & \graycell 1.15 $\times$ \\ \hline
\hline
\multicolumn{10}{|l|}{\textbf{~~GPT-4 Turbo}} \\ \hline
\multirow{2}{*}{CoT - Zero-Shot} & \xmark  & 70.48 & 82.04 &  86.34 & 84.91 & 69.78 & 26.81k & 6.79 &  -- \\
 & \graycell \cmark & \graycell 70.08 & \graycell 82.25 & \graycell 87.64 & \graycell 84.42 & \graycell 70.14 & \graycell 26.91k & \graycell 5.16 & \graycell 1.32 $\times$ \\ \hline
\multirow{2}{*}{CoT - Few-Shot} & \xmark & 72.89 & 84.87 &  83.75 & 97.29 & 72.15 & 28.49k & 6.76 &  -- \\
 & \graycell \cmark   & \graycell 72.16 & \graycell 85.48 & \graycell 82.95 & \graycell 99.72 & \graycell 72.72 & \graycell 28.92k & \graycell 5.09 & \graycell 1.33 $\times$ \\ \hline
\multirow{2}{*}{ReAct - Zero-Shot} & \xmark  & 74.30 & 85.80 & 88.49 & 94.52 & 72.18 & 30.51k & 6.67 & --  \\
 & \graycell \cmark  & \graycell 74.70 & \graycell 85.46  & \graycell 89.27 & \graycell 92.89 & \graycell 71.85 & \graycell 30.45k & \graycell 5.70 & \graycell 1.17 $\times$ \\ \hline
\multirow{2}{*}{ReAct - Few-Shot} & \xmark & 76.71 & 85.67 & 64.49 & 98.95 & 74.23 & 36.62k & 6.71 & --  \\
 & \graycell \cmark  & \graycell 76.28 & \graycell 85.46  & \graycell 65.17 & \graycell 99.50 & \graycell 74.13 & \graycell 36.68k & \graycell 5.72 & \graycell 1.17 $\times$ \\
\hline
\end{tabular}
\label{tab:results}
\end{center}
\end{table*}

\begin{table*}[htbp]
\caption{Zero-shot CoT (GPT-3.5 Turbo) runtime shows that overall latency reduction is highly dependent on data reuse rates. At high reuse, we observe only slight variability among different cache policies.}
\begin{center}
\begin{tabular}{|c||c||c|c|c|c|c||c|c|c|}
\hline
Cache Policy & No Cache &  \multicolumn{5}{c||}{LRU} & LFU & RR & FIFO \\ \hline 
Data Reuse Rate & -- & 0\% & 20\% & 40\% & 60\% & \graycell 80\% & 80\% & 80\% & 80\% \\ \hline
Avg Time/Task (s) $\downarrow$ &  5.81 & 5.81 & 5.84 & 5.62 & 5.03 & \graycell{4.92} & 5.16 & 5.36 & 5.25 \\
\hline
\end{tabular}
\label{tab:perf_vs_reuse}
\end{center}
\end{table*}

\section{Experimental Setup}
\label{sec:setup}

We use GeoLLM-Engine~\cite{singh2024geollmengine}, a large-scale parameterizable LLM engine for geospatial tasks. Designed to capture agentic performance across hundreds of tools, the platform is equipped with long-horizon multi-tool LLM operations that require frequent data retrieval and filtering, a comprehensive suite of open-source APIs, interactive map UI, RAG, and data retrieval tools with over 1.1 million satellite images.

\textbf{Benchmark}. We expand the GeoLLM-Engine \texttt{sampler} to obtain variants of the \texttt{GeoLLM-Engine-1k} dataset. Specifically, we extend the sampling-rate parameters and we incorporate rates that control the likelihood of data reuse. We selectively sample prompts with an 80\% probability of requiring data already present in the cache, constructing a test dataset of 1,000 multi-step prompts (with an overall set of approximately 50,000 tool calls). Additionally, we prepare a mini 500 query set for ablations. Last, we use the \texttt{model-checker} module to verify the functional correctness of the generated tasks.

\textbf{Metrics}. For agent performance, we adhere to established evaluation practices~\cite{zhuang2024toolqa, singh2024geoqa}, measuring the \textit{Success Rate} (proportion of tasks successfully completed), the \textit{Correctness Ratio} (proportion of correct tool calls, since an erroneous tool might not affect successful task completion), and the \textit{ROUGE-L} score. We also report performance on the underlying remote sensing tasks, with F1 and recall for object detection and land coverage classification (LCC), respectively, and ROUGE for visual question answering (VQA)~\cite{singh2024geollmengine}. 

To evaluate cache effectiveness, we report GPT-hits (\textit{i.e.}, the LLM correctly utilizes the cache over main memory). We also track the average number of tokens and time per task, with an expectation that higher cache reuse (being 5-10$\times$ faster than main memory access) will result in reduced overall API completion times. To capture latency, we follow~\cite{singh2024llmcompiler} by maintaining a running average per tool operation, discarding any outliers beyond two standard deviations from the mean. To avoid congestion and ensure accurate endpoint response times, we deploy hundreds of GPT instances specifically for this evaluation, isolated from production traffic.

\begin{table*}[htbp]
\caption{GPT-driven cache operations produce performance metrics and latency very similar to programmatic implementation of caching, demonstrating GPT's ability to successfully execute system optimization tasks.}
\begin{center}
\begin{tabular}{|c||c|c||c|c|c|c|c|c|c|c|c|}
\hline
\multirow{2}{*}{Model} & Cache & Policy & Cache Hit & Success  & Correctness  & Obj. Det & LCC  & VQA  & Avg Tokens & Avg Time \\
 & Read & Imp. & Rate (\%) $\uparrow$ & Rt (\%) $\uparrow$ & Rt (\%) $\uparrow$ & F1 (\%) $\uparrow$ & R (\%) $\uparrow$ & Rouge-L $\uparrow$ & /Task $\downarrow$ & / Task (s) $\downarrow$  \\
\hline
& Python & Python & - & 72.49 & 85.40 &  85.11 & 99.46 & 72.64 & 28.76k & 5.07 \\
\textbf{GPT-4 Turbo} & GPT-4 & Python & 96.59  & 72.16 & 85.41  & 83.00 & 98.69 & 72.35 & 28.73k & 5.11 \\
CoT - Few-Shot& Python & GPT-4 & 97.73 & 72.29 & 84.75 & 82.79 & 99.59 & 72.09 & 28.64k & 5.09 \\
  & \graycell GPT-4 & \graycell GPT-4 & \graycell 96.16 & \graycell 72.16 & \graycell 85.48  & \graycell 82.95 & \graycell 99.72 & \graycell 72.72 & \graycell 28.92k & \graycell 5.09 \\
\hline
\end{tabular}
\label{tab:results_vs_python}
\end{center}
\end{table*}

\section{Results}

LLM-dCache improves task-completion times across different configurations -- GPT-4 and GPT-3.5, with Chain-of-Thought and ReAct, in both few-shot and zero-shot scenarios -- by 1.24$\times$ on average (Table~\ref{tab:results}). Caching does not degrade the quality of output and functionality of the agent, as agent metrics are within established variance~\cite{singh2024llmcompiler}. Overall, we notice that gains primarily depend on dataset reusability patterns, not the choice of model or prompting strategy. 

To corroborate this observation, we conduct an ablation with multiple \texttt{mini-val} subsets, each containing 500 queries but with varying reusability rates. Table~\ref{tab:perf_vs_reuse} (top) shows higher reusability rates correlate with greater latency savings. LRU, LFU, RR, and FIFO produce no clear latency differences.

We aim to position our exploration within a broader shift towards empowering LLMs with system-level optimization decisions. To this end, we make the deliberate choice of treating cache operations as prompt-based GPT tools (\textit{e.g.}, explaining the LRU scheme via prompts) instead of a direct programmatic implementation of the logic. In support of this, our ablation in Table~\ref{tab:results_vs_python} compares programmatic cache operations with those driven by GPT. We find that all GPT-driven variants closely match the fully programmatic approach, which could be considered an upper-bound in terms of effectiveness and reliability, with cache ``hit'' rates consistently around 97\% and similar latency. This demonstrates the versatility and potential of LLM-guided cache management in lieu of traditional programmatic solutions. Our hope is that this perspective will motivate work for integrating LLMs into other system design optimizations~\cite{stamoulis2019single}, from execution at the edge~\cite{erdogan2024tinyagent} to energy/power optimizations and thermal management~\cite{marculescu2018hardware}.

\textbf{Limitations and future work}. Our study focuses on agentic performance and average latency for cloud-first environments with extensive use of cloud endpoints. It is meaningful to include more system performance metrics, such as energy and power consumption. To this end, we will explore GPT alternatives that can be run locally, such as Llama-3 and Phi-3.5. Given that our approach implements cache operations as callable API tools, we should be able to seamlessly incorporate this with other non-GPT tool-augmented agents across different computational environments. Last, we plan to extend our evaluation beyond the geospatial domain to a wider range of orthogonal tasks also considered in recent system-level LLM optimization papers~\cite{kim2024llmcompiler,fore2024geckopt}.

\section{Conclusion}

In this paper, we introduced LLM-dCache, a framework designed to optimize LLM data access patterns through a cache mechanism treated as callable API tools. By allowing LLMs to autonomously manage cache operations, we integrated caching with existing function-calling mechanisms, enabling improvements in system efficiency across various models and prompting techniques. Our work underscores the potential of leveraging LLMs for system-level optimizations in complex data-intensive environments. 

\section*{Acknowledgements}

This work is supported in part by grant NSF CCF 2324854.

\bibliographystyle{IEEEtran}
\bibliography{base}

\end{document}